\documentclass[12pt]{article}

\usepackage{amsmath}
\usepackage{amssymb}
\usepackage{graphicx}
\usepackage{array, longtable, multirow}
\usepackage[text={160mm,240mm},centering]{geometry}
\usepackage{float}

\newcommand{\gx}{$\tilde{\rm x}$}

\begin{document}

\begin{flushright}
USTC-ICTS-14-16\\
\end{flushright}

\vspace{20mm}

\begin{center}
{\Large \textbf{A classification of long-range interactions\\
 between two stacks of $p$ \& $p'$-branes}}

\vspace{6mm}

Jun Ouyang\footnote{yangjun@mail.ustc.edu.cn} and Chao Wu\footnote{wuchao86@mail.ustc.edu.cn}

\vspace{6mm}

\emph{Interdisciplinary Center for Theoretical Study\\
University of Science and Technology of China, Hefei, Anhui 230026, China}
\end{center}

\vspace{10mm}

\begin{abstract}
  We generalize the computations of the long-range interactions between two parallel stacks of branes in \cite{Wu and Wang, Prof. Lu's lecture notes} to various cases when two stacks of branes are not placed parallel to each other. We classify the nature of interaction (repulsive or attractive) for each special case and this
  classification can be used to justify the nature of long-range interaction between two complicated brane systems such as brane bound states. We will provide
  explicit examples in this paper to demonstrate this.
\end{abstract}

\newpage

\section{Introduction}

Strings and branes are basic dynamical objects in string/M theory \cite{string solitons}, like quarks and leptons in particle physics standard model. As such, understanding their interactions is always important, especially for revealing the non-perturbative nature of this theory, since except for the fundamental
strings, all the other brane objects are non-perturbative with respect to the strings.  In general, due to the lack of a complete formulation of this theory,
we don't have a full-fledged knowledge of interactions between any two (same or different type of) branes. However, we do have a knowledge of the leading-order interaction, i.e., the long-range one, between any two such branes. This is because we know the low-energy effective theories for both the bulk propagating modes
 (the corresponding supergravity theories) and the branes (the brane actions). From the former, we can determine the propagators for the bulk massless modes, responsible for the long-range interactions, and from the latter, we can determine their couplings with the branes, as discussed in detail in \cite{Wu and Wang, Prof. Lu's lecture notes}.

 Each of such basic branes has a tension, therefore a mass. So any two branes, whether the same or different type, must have a long-range gravitational interaction between the two. They carry also a charge, whose type depends on that of branes considered. For fundamental strings or the so-called NSNS 5-branes,
 they carry the so-called NS-NS charge, while for D$p$-branes, each of them carries a so-called R-R charge. Just like the electric charge, any two branes of the same type
 experience also a charge interaction and sometimes even for branes of different type, such as the one between D0 and D8 branes (there exists also a charge interaction between the two) \cite{Danielsson:1997wq, Bergman:1997gf,Lu and Wu}. These basic branes are actually stable, usually called BPS-branes, since in proper units
 their mass and charge are equal. When two such branes of the same type are placed parallel to each other at a separation, there exists no net force acting between the two since gravitational attractive force due to their tension cancels the repulsive one between the two due to their charge \cite{Wu and Wang}.

 As mentioned above, obtaining the full-fledged brane interaction is in general difficult or impossible for the time being.  Our knowledge about the corresponding long-range interaction can still be useful in many ways. For examples, with this information, we can determine whether any two branes can form bound orbit or bound state, and study the leading-order scattering process or amplitude. With the knowledge about the nature of the long-range interaction being attractive, zero or repulsive, we have at least partial knowledge to determine whether the underlying system has a possibility to form a stable configuration.

 Unlike point particles in field theory, the branes in string/M-theory can have different spatial dimensionalities and the interaction nature depends on the relative
 orientation between two such branes.  In this paper, we will give a classification of the long-range interaction nature between any two simple BPS branes of the same or different type  in spacetime $D = 10$ and $D = 11$,  placed with various orientations specified later. This classification has its use or advantage in that once such interaction nature for the simple BPS branes are known, we can use this knowledge to determine the nature of long-range interaction for complicated brane systems such
 as brane bound states, without the need of computations. For example, if we table the classification, we can simply read the interactions between the constituent branes in the two brane bound states, from the table and often we can determine the nature of the underlying interaction. We will provide explicit examples for this purpose.

In \cite{Wu and Wang}, the authors calculate the long-range Coulomb-type interaction between two stacks of $p$-branes, placed parallel at a separation, via the effective field theory computations. Similar calculations are made in \cite{Prof. Lu's lecture notes} for the cases when two stacks of branes of different type,  are placed parallel to each other. We in this paper consider the various cases when the two stacks of branes, being the same or different type, are placed non-parallel to each other, in $D = 10$ and 11. We will classify the interaction nature for the cases given in \cite{Wu and Wang, Prof. Lu's lecture notes} and computed in this paper.

In our computations, we use the integer $q$, the common dimensions that the two branes share, to describe the extent of non-parallel. For example, when $q = 0$, the $p$-branes and $p'$-branes, placed at a separation, are actually orthogonal to each other in space.  Combining with the results given in \cite{Wu and Wang} and in \cite{Prof. Lu's lecture notes}, we find that the net interaction depends on the three integers $q$, $p$ and $p'$, in addition to the spacetime dimensions $D$. In the classification, we find that the interaction nature is closely related to the value of integer $q$.

This paper is organized as follows: In section 2, we have a review on how to get the bulk propagators and couplings from the bulk action and effective brane world-volume action. In section 3, we review the calculation of the long-range interactions in \cite{Prof. Lu's lecture notes} between two stacks of parallel branes at a separation in diverse dimensions, both of the same and different kinds. The long-range interaction between two stacks of non-parallel branes are calculated in section 4. In section 5, using the results in section 3 and 4, we have a classification about the nature of the interaction for parallel and non-parallel branes as forms of tables. Applications of these results will be mentioned in section 6, about determining the nature of total interactions qualitatively between two stacks of D-branes with world-volume abelian or non-abelian flux. Section 7 is a brief summary.

\section{The set up}

The long-range interaction between any two stacks of static BPS $p$- and $p'$-branes, parallel or non-parallel, in a separation, are due to the exchanges of relevant bulk bosonic propagating modes. For computing this interaction, we need to have the relevant  bulk bosonic field propagators and  the corresponding couplings of these fields with p-branes under consideration, following \cite{Wu and Wang, Prof. Lu's lecture notes}. For this, we consider the relevant bosonic part of D-dimensional supergravity action in $p$-brane frame as \cite{Duff1994}
\begin{align}\label{sugra_action_pbraneframe}
  I_{D}(d)=&\frac{1}{2\kappa_{D}^{2}}\int d^{D}x\sqrt{-G}e^{-\frac{(D-2)\alpha(d)\Phi}{2d}}\left[R-\frac{1}{2}\Big(1-\frac{\alpha^{2}(d)(D-1)(D-2)}{2d^{2}}\Big)(\nabla\Phi)^{2} \right. \nonumber \\
                  &\left.-\frac{1}{2}|F_{d+1}|^{2}\right].
\end{align}
In the above, $G_{MN}$ is the spacetime metric in $p$-brane frame,  $R$ the Ricci scalar, $\Phi$ the dilaton with its vacuum expectation value $\Phi_0$, and $F_{d+1}$ is the field strength of $d=p+1$ form field $A_d$. $\alpha(d)$ is the dilaton coupling and is given as
\begin{equation}\label{eq: alpha d 2}
  \alpha^2(d)=4-\frac{2(p+1)(D-p-3)}{D-2}.
\end{equation}
Since our focus on the long-range interaction is in the field theory limit,  it is therefore proper to express the above action in Einstein or canonical frame. This can be achieved by making the following rescaling of metric
\begin{equation}\label{eq: brane frame to canonical frame}
  G_{MN}=e^{\frac{\alpha(d)\phi}{p+1}}g_{MN},
\end{equation}
where $\phi=\Phi-\Phi_0$.
Thus we have (\ref{sugra_action_pbraneframe}) in canonical frame:
\begin{equation}\label{eq: bulk action in cononical frame}
   I_{D}(d)=\frac{1}{2\kappa^{2}}\int d^{D}x\sqrt{-g}\left[R-\frac{1}{2}(\nabla \phi)^{2}-\frac{1}{2}e^{-\alpha(d)\phi}|F_{d+1}|^{2}\right],
\end{equation}
where we define the physical gravitational coupling $2\kappa^2=2g_b^2\kappa_D^2$ with the dimensionless parameter $g_b= e^{\frac{(D-2)\alpha(d)}{4 (p+1)}\Phi_0}$. When we set $D=10$ and $p=1$, $g_b$ becomes the string coupling $g_s$.\\

We now consider small fluctuations of fields with respect to the vacuum, i.e.,  $g_{MN}=\eta_{MN}+h_{MN}$ with $\eta_{MN}$ the flat Minkowski metric, $\Phi = \Phi_0 + \phi$ and $A_d = 0 + A_d$. We also choose the usual harmonic gauge for $h_{MN}$:  $\partial_{P}h^{P}_{N}-\frac{1}{2}\partial_{N}h=0$. Then the action becomes:
\begin{equation}\label{eq: leading order bulk action}
  I_D(d) = \frac{1}{2 \kappa ^2}\int d^D x \left[-\frac{1}{4}\nabla h^{MN}\nabla h_{MN} + \frac{1}{8}(\nabla h)^2-\frac{1}{2}\left(\nabla\phi\right)^2-\frac{1}{2}|F_{d+1}|^2\right],
\end{equation}
where we keep only the lowest order terms. In order to have  the propagators and the couplings with the correct normalization, we need to make the following rescalings of fields as:
\begin{equation}\label{eq: rescaling}
  h_{MN} \rightarrow 2\kappa h_{MN},~~~\phi \rightarrow \sqrt2\kappa\phi,~~~A_{p+1} \rightarrow \sqrt2\kappa A_{p+1},
\end{equation}
then we have
\begin{equation}\label{gaction}
I_{D}(d)=\int d^{D}x\left[-\frac{1}{2}\nabla h^{MN}\nabla h_{MN}+\frac{1}{4}(\nabla h)^{2}-\frac{1}{2}(\nabla \phi)^{2}-\frac{1}{2}| F_{d+1}|^{2}\right],
\end{equation}
from which we can read the propagators for graviton, dilaton and d-form field, in momentum space, as
\begin{align}\label{eq: propagators}
\underbrace{h_{MN}h_{M'N'}}&=\frac{1}{k_{\perp}^{2}}\left[\frac{1}{2}\eta_{MM'}\eta_{NN'}+\frac{1}{2}\eta_{MN'}\eta_{NM'}-\frac{1}{D-2}
\eta_{MN}\eta_{M'N'}\right], \\
\underbrace{\phi \phi}&=\frac{1}{k_{\perp}^{2}}, \\
\underbrace{A_{01\ldots p}A_{01\ldots p'}}&=-\frac{1}{k_{\perp}^{2}}\delta_{p,p'}.
\end{align}
In the above, $k_{\perp}$ is the spatial momentum along the directions perpendicular to the brane.

The bosonic part of the world volume action for a $p$-brane is
\begin{equation}
  S_{d}=-T_{p}\int d^{p+1}\sigma \sqrt{-G}+T_{p}\int A_{p+1},
\end{equation}
where $p$-brane frame metric $G_{\alpha\beta}$\footnote{$\alpha$, $\beta$, $\gamma\cdots$ are the world volume indexes} and the $(p+1)$ form field $A_{p+1}$ are the pull back of the corresponding bulk fields to the world volume, and $T_p$ is the brane tension. Using equation (\ref{eq: brane frame to canonical frame}), we can express the above in the canonical frame as
\begin{equation}
  S_{d}=-T_{p}\int d^{p+1}\sigma e^{\frac{\alpha(d)\phi}{2}}\sqrt{-g}-\frac{T_{p}}{(p+1)!}\int d^{p+1}\sigma \epsilon^{\alpha_{0}\cdots\alpha_{p}} A_{\alpha_{0}\cdots\alpha_{p}},
\end{equation}
where $g_{\alpha\beta}$ is the pull back of canonical metric to the world volume and $\epsilon^{\alpha_{0}\cdots\alpha_{p}}$ is totally antisymmetric with respect to its indices with $\epsilon^{0\cdots p} =  1$. With the same small fluctuations of background,  we expand the above action to the leading order and have
\begin{equation}\label{eq: leading order brane action}
  S_{d}=-T_{p}\int d^{p+1}\sigma \left(1+\frac{\alpha(d)\phi}{2}+\frac{1}{2}\eta^{\alpha\beta}h_{\alpha\beta}\right)-\frac{T_{p}}{(p+1)!}\int d^{p+1}\sigma \epsilon^{\alpha_{0}\ldots\alpha_{p}} A_{\alpha_{0}\ldots\alpha_{p}},
\end{equation}
 With the scalings (\ref{eq: rescaling}) and noticing that $\phi$, $h_{\alpha\beta}$ and $A_{\alpha_{0}\ldots\alpha_{p}}$ depend only on the coordinates transverse to the brane, we then have
\begin{align}
S_{d}=&-T_{p}V_{p+1}-T_{p}V_{p+1}\kappa \eta^{\alpha \beta}h_{\alpha \beta}-\frac{1}{\sqrt{2}}T_{p}V_{p+1}\kappa \alpha (d)\phi \nonumber \\
            &+\frac{\sqrt{2}\kappa T_{P}}{(p+1)!}V_{p+1}\epsilon^{\alpha_{0}\alpha_{1}\ldots \alpha_{p}}A_{\alpha_{0}\alpha_{1}\ldots \alpha_{p}}.
\end{align}
So the couplings of the brane with bulk fluctuations can be read,respectively, as
\begin{align}\label{eq: couplings}
J_{h}^{(i)}&=-c_{p}V_{p+1}N_{i}\eta^{\alpha \beta}h_{\alpha \beta},\nonumber \\
J_{\phi}^{(i)}&=-\frac{1}{\sqrt{2}}c_{p}V_{p+1}N_{i}\alpha(d)\phi, \nonumber \\
J_{A_{p+1}}^{(i)}&=\frac{\sqrt{2}c_{p}V_{p+1}N_{i}}{(p+1)!}\epsilon^{\alpha_{0}\alpha_{1}\ldots \alpha_{p}}A_{\alpha_{0}\alpha_{1}\ldots \alpha_{p}}.
\end{align}
In the above, we have defined $c_{p} \equiv T_{p}\kappa$ and introduce an extra factor $N_{i}$ to count the multiplicity of  the branes. $i=1,2$ refers to the two stacks of branes, respectively. $V_{p+1}$ is the volume of $p$-brane's world volume.

\section{The long-range interactions between two parallel stacks of $p$ \& $p'$-branes}\label{parallel brane interaction}

In this section we will use the propagators and couplings obtained from the previous section to calculate the long-range interactions in momentum space\footnote{ The interaction in coordinate space can be obtained simply by a Fourier transformation following \cite{lu ning wei xu: interaction between}.} between two parallel stacks of $p$-branes and $p'$-branes, due to the exchange of massless modes in spacetime $D=10$ and $D=11$.

The interaction due to the exchange of graviton is
\begin{equation}
U_{h}=\underbrace{J_{h}^{(1)}J_{h}^{(2)}}=c_{p}c_{p'}V_{p+1}V_{p'+1}N_{1}N_{2}\eta ^{\alpha \beta}\eta ^{\gamma \delta}\underbrace{h_{\alpha \beta}h_{\gamma \delta}},
\end{equation}
Using the propagator in (\ref{eq: propagators}), we have
\begin{equation}\label{eq: gravitation interaction}
U_{h}=c_{p}c_{p'}V_{p+1}V_{p'+1}N_{1}N_{2}\frac{1}{k^2_{\perp}}\left[(p'+1)-\frac{1}{D-2}(p+1)(p'+1)\right].
\end{equation}
Using (\ref{eq: propagators}) and (\ref{eq: couplings}), the interaction due to the exchange of dilaton is
\begin{equation}\label{eq: dilaton interaction}
U_{\phi}=\underbrace{J_{\phi}^{(1)}J_{\phi}^{(2)}}=\frac{1}{2k^2_\bot}c_{p}c_{p'}V_{p+1}V_{p'+1}N_{1}N_{2}\alpha(d)\alpha(d'),
\end{equation}
where
\begin{equation}\label{alphad}
\alpha(d)=\begin{cases}
                \frac{3-p}{2}& \qquad {\text{for NS-NS p-brane}};\\
                \frac{p-3}{2}& \qquad {\text{for $Dp$ brane}}.
                \end{cases}
\end{equation}
and in general
\begin{equation}\label{alpha2}
  \alpha^2(d)=4-\frac{2(p+1)(D-p-3)}{D-2}.
\end{equation}
The equation (\ref{eq: dilaton interaction}) indicates that for different type of branes, the interaction due to the exchange of dilaton depends not only on the dimensionality of each stack of branes under consideration, but also on the dilaton coupling $\alpha(d)$.

The interaction due to the exchange of the ($p+1$)-form potential $A_{01\ldots p}$ also can be calculated as
\begin{equation}\label{eq: charge interaction}
U_{A_{p+1}/A_{p'+1}}=\underbrace{J_{A_{p+1}}^{(1)}J_{A_{p'+1}}^{(2)}}=-2c_{p}c_{p'}V_{p+1}V_{p'+1}N_{1}N_{2} \frac{1}{k_{\perp}^{2}}\delta_{p,p'}.
\end{equation}
To have this contribution non-vanishing, the $p$-branes and $p'$-branes must be the same type or be mutually branes and anti-branes with $p = p'$ and be placed parallel to each other. In the former case, the interaction is repulsive while for the latter it is attractive and the overall sign on the right side of above equation is positive instead. For now, we will exclude the case of branes and antibranes from consideration.

So the total interaction is
\begin{align}\label{eq: total interaction_p brane and p' brane}
U&=U_{h}+U_{\phi}+U_{A_{p+1}/A_{p'+1}} \nonumber \\
   &=c_{p}c_{p'}V_{p+1}V_{p'+1}N_{1}N_{2} \frac{1}{k_{\perp}^{2}}\left[\frac{(p'+1)(D-2)-(p+1)(p'+1)}{D-2}+\frac{\alpha(d)\alpha(d')}{2}-2\delta_{p,p'}\right] \nonumber \\
   &=\frac{\eta}{k_{\perp}^{2}}c_{p}c_{p'}V_{p+1}V_{p'+1}N_{1}N_{2}.
\end{align}
where\footnote{If the two sets of branes are mutually branes and anti-branes, the term $- 2 \delta_{p,p'}$ in the square bracket should be replaced by $2 \delta_{p, p'}$ instead.}
\begin{equation}\label{etaparallel}
\eta=\frac{(p'+1)(D-2)-(p+1)(p'+1)}{D-2}+\frac{\alpha(d)\alpha(d')}{2}-2\delta_{p,p'}.
\end{equation}
Because the factors in front of the bracket in the second line of (\ref{eq: total interaction_p brane and p' brane}) are all positive, so we can use $\eta$ to determine the properties of the total interaction. If $\eta>0$, the interaction is attractive while $\eta<0$ it is repulsive. When $\eta=0$, there is no interaction between the two stacks of $p$ and $p'$-branes.

For M-theory, we have M2-branes and M5-branes but there is no dilaton. The relevant bulk action is the bosonic part of $D=11$ supergravity
\begin{equation}
   I_{11}=\frac{1}{2\kappa^{2}}\int d^{11}x\sqrt{-g}\left[R-\frac{1}{2}|F_{4}|^{2}\right]-\frac{1}{12\kappa^{2}}\int F_{4}\wedge F_{4}\wedge A_{3},
\end{equation}
where $F_{4}$ is the 4-form field strength of the 3-form potential $A_{3}$. The second term is a higher order term, so has no contribution to the long-range interaction for small fluctuations. For this reason, the action is just a special case of (\ref{gaction}). Therefore, the general result in (\ref{eq: total interaction_p brane and p' brane}) is also applicable here for $p =2$ or $5$. One can check that using (\ref{eq: alpha d 2}), we have indeed $\alpha (3) = \alpha (6) = 0$ when $D = 11$.

For simplicity, we assume for now that $p\geq p'$, so the parameters satisfy $0\leq p' \leq p \leq D-2$  since we need to have at least one spatial dimension transverse to both stacks of branes under consideration (For the interaction to have a good behavior in coordinate space, we need to extend the upper bound to $D-4$ instead).

\subsection{$D=11$}
For M-theory, there are M5-branes and M2-branes but there is no dilaton. Therefore $U_{\phi}=0$ and $\eta=\frac{(8-p)(p'+1)}{9}-2\delta_{p,p'}$. We have the following three subcases to consider:

\begin{itemize}
\item
M5- and M2-branes
\\In this case $p=5, p'=2$, so $\eta=1$. The interaction between M2- and M5-branes is attractive because there is only gravitational interaction between them.

\item
M5- and M5-branes
\\In this case $p = p' =5$, so $\eta=0$. The contributions from graviton and form potential exchanges cancel each other, so there is no interaction between M5- and M5-branes.

\item
M2- and M2-branes
\\In this case $p=p'=2$, so $\eta=0$. Like the M5- and M5-branes case, there is no interaction between M2- and M2-branes too.

\end{itemize}

\subsection{$D=10$}
\subsubsection{$p>p'$}
At first, we consider the situations for $p \neq p'$ and assume that $p>p'$. Equation (\ref{eq: charge interaction}) indicates that there is no contribution from the $(p+1)$-form fields, so the total interaction is determined by the contributions from graviton and dilaton exchanges.
\begin{itemize}
  \item
  D$p$ and D$p'$-branes
  \\In this case $\alpha (d)=\frac{p-3}{2}$, $\alpha (d')=\frac{p'-3}{2}$, and $p-p'= {\rm even} =2,4,6,8$, so $\eta=\frac{p'-p+4}{2}$.
  \begin{itemize}
    \item
    If $p-p'=2$, $\eta=1$, so the interaction between D$p$ and D$(p-2)$-branes is attractive.
    \item
    If $p-p'=4$, $\eta=0$. The contributions from graviton and dilaton exchanges cancel each other, so there is no interaction between D$p$ and D$(p-4)$-branes.
    \item
    If $p-p'=6$, $\eta<0$, so the interaction between D0 and D6-branes, or between D1 and D7-branes is repulsive\footnote{The interaction between D0 and D8-branes is more complicated. Because there is an additional duality relation $A_{0}=-A_{01\dots 8}$, thus can lead to an extra attractive coupling between the one-form potential $A_{1}$ and the nine-form potential $A_{9}$, and the coupling can be interpreted as arising from the half-string creation \cite{Danielsson:1997wq,Bergman:1997gf,Kitao1998,Billo1998} between D0-branes and D8-branes. The detail analysis can be found in \cite{Lu and Wu} and \cite{Wu and Wang}. We will exclude this case from consideration in this paper. In other words, we limit ourselves to consider only $p - p' \le 6$.}. That means the repulsive interaction due to the dilaton exchange overtakes that due to the graviton exchange.
  \end{itemize}

  \item
  D$p$-branes and F-strings ($p>1$)
  \\In this case, $\alpha (d)=\frac{p-3}{2}$ and $\alpha (d')=\frac{3-1}{2}=1$, thus we have $\eta=1$. So the interaction between the two stacks of D$p$-branes and fundamental strings is attractive.

  \item
  D0-branes and F-strings
  \\In this case, $p=1$, $\alpha (d)=\frac{3-1}{2}=1$, and $p'=0$, $\alpha (d')=\frac{0-3}{2}=-3/2$, so $\eta=0$. There is no interaction between D0-branes and fundamental strings.

  \item
  D$n$ and NS5-branes
  \begin{itemize}
  \item
  If $n>5$
  \\In this case we should take $p'=5$ and $p=n$ in the formula of $\eta$ (\ref{etaparallel}), $\alpha (d')=\frac{3-5}{2}=-1$, and $\alpha (d)=\frac{n-3}{2}$, we can get $\eta=6-n\leq 0$. For $n=6$, the interaction between D6 and NS5-branes vanishes and this is consistent with that between D0 and F-strings since these two systems are related to each other by spacetime Hodge-duality. For $n>6$, the interaction between them is repulsive.
  \item
  If $n<5$
  \\In this case we should take $p=5$ and $p'=n$ in the formula of $\eta$ (\ref{etaparallel}). So we have $\alpha (d)=\frac{3-5}{2}=-1$, and $0\leq n<5$, $\alpha (d')=\frac{n-3}{2}$. So we have now $\eta=1$, it means that the interaction between D$n$-branes and NS5-branes is attractive.
  \end{itemize}

  \item
  F-strings and NS5-branes
  \\In this case $p'=1$, $\alpha (d')=\frac{3-1}{2}=1$, and $p=5$, $\alpha (d)=\frac{3-5}{2}=-1$. So $\eta=0$, it means that there is no interaction between two stacks of fundamental strings and NS5-branes, and this is consistent with their S-dual result between D1 and D5-branes in the Type IIB theory.

  \item
  For the special case $p' = \tilde p$ with $\tilde{p}$-brane as the Hodge-dual of $p$-brane
  \\The total spacetime dimension $D=d+\tilde{d}+2=p+\tilde{p}+4$ and $\alpha (d')=\alpha (\tilde{d})=-\alpha (d)$. So we have $p'=D-4-p<p$,  which gives $p\geq \frac{D-4}{2}$. The interaction between $p$-branes and its Hodge-dual $\tilde{p}$-branes can be explicitly given in any $D$ using the formula of $\eta$ in (\ref{etaparallel}) and the formula for $\alpha^{2}(d)$ in (\ref{alpha2}). Here we mention a few examples which have been discussed above already. In $D=10$, we have D4-D2,D5-D1,D6-D0 and NS5-F system and in $D=11$ we have the M5-M2 system.

\end{itemize}

\subsubsection{$p=p'$}
Secondly, we will consider the case for $p=p'$. There are two different subcases:
\begin{itemize}
  \item
  The same type of $p$ and $p'$-branes: Dp-Dp,F1-F1,NS5-NS5
  \\ We have now nonzero and repulsive interaction from the $(p+1)$-form potential exchange. Using (\ref{eq: alpha d 2}), we have
  \begin{equation}
  \eta=\frac{(7-p)(p+1)}{8}+\frac{\alpha ^{2}(d)}{2}-2=\frac{(7-p)(p+1)}{8}+2-\frac {(p+1)(7-p)}{8}-2=0.
  \end{equation}
  It means that for two stacks of $p$-branes with the same type, the contributions from graviton, dilaton, and ($p+1$)-form potential exchanges cancel exactly. It is the so-called ``no-force" condition for BPS branes, and this configuration preserves $1/2$ of spacetime supersymmetries \cite{Wu and Wang}.

  \item
  The different type of $p$ and $p'$-branes: F1-D1,D5-NS5
  \\Because $p$-branes and $p'$-branes are two different types of branes, there is no interaction from the $(p+1)$-form potential between them although their dimensionalities are equal. Therefore the total interaction is
  \begin{equation}
  \eta=\frac{(7-p)(p+1)}{8}+\frac{(p-3)(3-p)}{8}=\frac{-p^{2}+6p-1}{4}.
  \end{equation}
  When $p=1$ or 5, we both have $\eta=1$. So the interaction between fundamental strings and D-strings, or between D5 and NS5-branes is attractive, respectively.
\end{itemize}

\section{The long-range interactions between non-parallel branes}\label{non-parallel brane}
In this section we consider the situations that the two stacks of $p$-branes and $p'$-branes are not placed parallel to each other. We still assume $p' \leq p$ and introduce an additional parameter $q$ to denote the commom dimensions that the two branes share. $q=0$ means that the two stacks of p-branes and $p'$-branes, placed at a separation, are actually orthogonal to each other. On the other hand, $q=p'$ means that the two stacks of p-branes and $p'$-branes are parallel to each other. This latter case corresponds to what we have discussed in the previous section and will be excluded in what follows. As before, for the purpose of calculating the long-range interaction, there is at least one spatial dimension transverse to both stacks of branes involved. In other words, we need $p < D-2$. So $q, p'$ and $p$ satisfy $\rm \max\{0, p+p'-(D-2)\}\leq q<p'\leq p<D-2$. The propagators and the couplings derived in section 2 continue to apply here. The interaction due to the graviton exchange is now
\begin{align}
  U_{h}&=\underbrace{J_{h}^{(1)}J_{h}^{(2)}}=c_{p}c_{p'}V_{p+1}V_{p'+1}N_{1}N_{2}\eta ^{\alpha \beta}\eta ^{\gamma \delta}\underbrace{h_{\alpha \beta}h_{\gamma
                  \delta}}\nonumber\\
            &=c_{p}c_{p'}V_{p+1}V_{p'+1}N_{1}N_{2}\frac{1}{k^2_{\perp}}\left[(q+1)-\frac{1}{D-2}(p+1)(p'+1)\right].
\end{align}
 The above indicates that this interaction depends on the dimensionality of each stack of branes involved, the spacetime dimension D and the common dimensions $q$ that the two stacks of branes share.

The interaction due to the exchange of dilaton remains the same as that in the parallel case and continues to be given by (\ref{eq: dilaton interaction}) with
$\alpha(d)$ given in (\ref{alphad}) when $D = 10$ and in (\ref{alpha2}) for a general $D$.

On the other hand, since the two stacks of $p$-branes and $p'$-branes are not placed parallel, we don't have contribution from the exchange of form-potentials even when $p = p'$. So we have the total interaction as
\begin{align}\label{eq: total interaction_non parallel_p and p' brane}
  U&=U_{h}+U_{\phi}+0 \nonumber \\
     &=c_{p}c_{p'}V_{p+1}V_{p'+1}N_{1}N_{2}\frac{1}{k_{\perp}^{2}}\left[\frac{(q+1)(D-2)-(p+1)(p'+1)}{D-2}+\frac{\alpha(d)\alpha(d')}{2}\right] \nonumber\\
     &=\frac{\eta}{k_{\perp}^{2}}c_{p}c_{p'}V_{p+1}V_{p'+1}N_{1}N_{2}.
\end{align}
where
\begin{equation}\label{etanonparallel}
\eta=\frac{(q+1)(D-2)-(p+1)(p'+1)}{D-2}+\frac{\alpha(d)\alpha(d')}{2}.
\end{equation}
By the same token,  we can use the sign of $\eta$ to determine whether the interaction is attractive or repulsive or vanishing.

\subsection{$D=11$}
In $M$-theory there are M5-branes and M2-branes. And $\alpha (3)=\alpha (6)=0$ because there is no dilaton, thus $U_{\phi}=0$, we have $\eta=\frac{9(q+1)-(p+1)(p'+1)}{9}$.

\begin{itemize}
\item
M5- and M2-branes
\\In this case $p=5, p'=2$, so $\eta=q-1$. The range of $q$ is $0\leq q<p'=2$, so $q=0, 1$. for $q=0$, the interaction between two stacks of M2-branes and M5-branes is repulsive. For $q=1$, there is no interaction between them and this is related to the D5/D1 system considered in the parallel case via dimensional reduction and T-dualities.

\item
M5- and M5-branes
\\In this case $p=p'=5$, so $\eta=q-3$. Because $\rm \max\{0, p+p'-(D-2)\}= 1 \leq q < p' = 5$, so $q=1, 2,3,4$. For $q=4$, the interaction between M5- and M5-branes is attractive. For $q=3$, there is  no interaction. For $q=1, 2$, the interaction becomes repulsive.

\item
M2- and M2-branes
\\In this case $p=p'=2$, so $\eta=q$ and can only be 0 or 1. For $q=1$, the interaction between M2- and M2-branes is attractive. For $q=0$, there is no interaction between them  and this system is related to the D0 $\&$ F system considered before via dimensional reduction and T-dualities. In both cases, there is no interaction and therefore this is a consistent result.
\end{itemize}

\subsection{$D=10$}
\subsubsection{$p>p'$}
In the following, we consider branes in string theory and we will not mention the D0-brane because it can't be orthogonal to any other kind of branes including itself. At first, we consider the situations for $p \neq p'$ and assume that $p>p'$.
\begin{itemize}
  \item
  D$p$ and D$p'$-branes
  \\In this case, $\alpha(d)=\frac{p-3}{2}$, $\alpha(d')=\frac{p'-3}{2}$ and $p-p'= {\rm even} =2,4,6$. So $\eta=q+\frac{4-p-p'}{2}$ with ${\rm max} \{0, p+p'-8\}\leq q<p'-1$.
\begin{itemize}
    \item
    If $p-p'=2$, $\eta=q+1-p'$. Thus when $q=p'-1$, there is no interaction between D$p$ and D$(p-2)$-branes. When ${\rm max}\{0, p+p'- 8\}\leq q<p'-1$, the interaction between them is repulsive.
    \item
    If $p-p'=4$, $\eta=q-p'<0$. So the interaction between two stacks of D$p$-branes and D$(p-4)$-branes is always repulsive.
    \item
    If $p-p'=6$, $\eta=q-p'-1<0$. So the total interaction between D$p$-branes and D$(p-6)$-branes is always repulsive.
  \end{itemize}

  \item
  D$p$-branes and F-strings ($p>1$)
  \\In this case, $\alpha (d)=\frac{p-3}{2}$ and $p'=1$, $\alpha (d')=\frac{3-1}{2}=1$, note that $q$ can only be $0$. Thus we have $\eta=0$, so there is no interaction between two stacks of D$p$-branes and fundamental strings.

  \item
  D$n$ and NS5-branes
  \begin{itemize}
  \item
  If $n>5$
  \\In this case we should take $p'=5$ and $p=n$ in the formula of $\eta$ (\ref{etanonparallel}), $\alpha (d')=\frac{3-5}{2}=-1$, and $\alpha (d)=\frac{n-3}{2}$. So $\eta=q+1-n<p'+1-n=6-n\leq 0$, so the interaction between D$n$ with $n > 5$ and NS5-branes is repulsive.
  \item
  If $n<5$
  \\In this case we should take $p=5$ and $p'=n$ in the formula of $\eta$ (\ref{etanonparallel}), $\alpha (d)=\frac{3-5}{2}=-1$, and $\alpha (d')=\frac{n-3}{2}$. So we have $\eta=q+1-n\leq 0$.
  If $q=n-1$, we have $\eta=0$, there is no interaction between D$n$ and NS5-branes. And If $q<n-1$, we have $\eta<0$, the interaction between them is repulsive.
  \end{itemize}

  \item
  F-strings and NS5-branes
  \\In this case $p=5$, $\alpha (d)=\frac{3-5}{2}=-1$, and $p'=1$,  $\alpha (d')=\frac{3-1}{2}=1$. Here $q = 0$.  So we have $\eta=-1$, thus the interaction between two stacks of fundamental strings and NS5-branes is repulsive, and this is consistent with their S-dual result between D1 and D5-branes discussed above in Type IIB theory.

  \item
  For the special case $p'=\tilde p$ with $\tilde{p}$-brane as the Hodge-dual of $p$-brane
  \\The total spacetime dimension $D=d+\tilde{d}+2=p+\tilde{p}+4$, and $\alpha (d')=\alpha (\tilde{d})=-\alpha (d)$. So we have $p'=D-4-p<p$, it means that $p\geq \frac{D-4}{2}$. Like the previous section, the interaction between $p$-branes and its Hodge-dual $\tilde{p}$-branes can be explicitly given in any $D$ using the formula of $\eta$ in (\ref{etanonparallel}) and the formula for $\alpha^{2}(d)$ in (\ref{alpha2}). Here we also mention a few examples which have been discussed above already. In $D=10$, we have D4-D2, D5-D1 and NS5-F systems, and in $D=11$ we have the M5-M2 system.
\end{itemize}

\subsubsection{$p=p'$}
Secondly, we will consider the cases for $p=p'$. There are two different subcases:
\begin{itemize}
  \item
  The same type of $p$ and $p'$-branes: Dp-Dp,~F1-F1,~NS5-NS5
  \\Using (\ref{eq: alpha d 2}), we have
  \begin{equation}
  \eta=\frac{8q+8-(p+1)^{2}}{8}+\frac{\alpha ^{2}(d)}{2}=q+\frac{8-(p+1)^{2}}{8}+2-\frac {(p+1)(7-p)}{8}=q+2-p.
  \end{equation}
  Therefore if $q=p-1$, $\eta =1>0$, the interaction between two stacks of $p$-branes is attractive. If $q=p-2$, $\eta=0$, there is no interaction. And if $q<p-2$, $\eta <0$, the interaction between them is repulsive.

  \item
  The different type of $p$ and $p'$-branes: F1-D1, D5-NS5
  \\In this case $\alpha(d)=\frac{p-3}{2}$, $\alpha(d')=\frac{3-p'}{2}=\frac{3-p}{2}$, so we have
  \begin{equation}
  \eta=\frac{8q+8-(p+1)^{2}}{8}+\frac{(p-3)(3-p)}{8}=q-\frac{(p-1)^{2}}{4}.
  \end{equation}
  For F1-D1, we have $p=1$ and $q=0$, so $\eta=0$, therefore there is no interaction between two stacks of fundamental strings and D1-branes. For D5-NS5, we have $p=5$, so $\eta=q-4$. Note that $\rm \max\{0, 2\}=2\leq q<p=5$, so $q=2,3,4$. Therefore if $q=4$, there is no interaction between D5 and NS5-branes, while if $q=2,3$, the interaction between them is repulsive.
\end{itemize}

\section{A classification of long-range interactions between two stacks of $p$ \& $p'$-branes}
In this section we will give a classification of the long-range interactions computed and discussed in  the previous sections, including both parallel and non-parallel cases. The purpose doing this is two-fold: 1) we can have a dictionary on the interactions for all simple p-branes, 2) with this, the nature of interaction for complicated systems such as brane bound states as mentioned in the introduction can be determined without the need of detail computations.

 Prior to this purpose, we would like first to use  T or/and S-duality to check consistencies for the results obtained in the previous two sections regarding the interactions for the parallel and non-parallel cases against the known results. We have already discussed some examples in the previous sections. We know that a T-duality interchanges the Neumann and Dirichlet boundary conditions but does not change the nature of interaction. A T-duality in a direction tangent to a D$p$-brane converts  it to a D$(p-1)$-brane, while in a direction orthogonal to it turns it into a D$(p+1)$-brane \cite{Clifford D-Brane}. For example, we can take a T-duality in a direction tangent to D$p$-branes once in a time for p-times and convert them to D0-branes.  As a more illustration, we consider the long-range interaction between non-parallel D1- and D5-branes given in section 4. This one is repulsive. Now we take a T-duality in the direction tangent to D1-branes which is orthogonal to the D5-branes. We then turn D1 to D0 while at the same time turn the D5 to D6 branes. In other words, we convert the D1 and D5 into an interacting system of D0 and D6, whose long range interaction is known to be repulsive, therefore a consistent check that the long-range interaction remains the same before and after a T-duality. Others can be similarly discussed.

 Now we summarize the results obtained in the previous two sections in spacetime $D=11$ and $D=10$ in Table 1 and Table 2, respectively. A more detail results in Table 3 show the nature of long-rang interactions between two stacks of non-parallel branes sharing various allowed common dimensions $q$ in $D=10$. In the tables, we assume $p' \leq p$ and denote the attractive and repulsive interactions using the sign $``+"$ or $``-"$ and the vanishing one with ``0", respectively. Also we use the symbol $``\diagup"$ to stand for the non-existence of a case. In Table 3, we list only the $p \ge p'$ cases and it should be understood that the $p' > p$ cases are symmetric to those of $p > p'$, which we don't show in the table.

\begin{table}[H]\centering
 \caption{The nature of long-range interaction between two stacks of parallel or non-parallel p $\&$ $p'$-branes in $D = 11$}
\begin{tabular}{|c|c|c|}
\hline
$D=11$         & Parallel       & Non-parallel \\
\hline
               & \quad                     & \,$q=1: +$ \\
 M2-M2       & 0                    & $q=0: 0$ \\

\hline
               &\quad                      & $q=1: 0$ \\
 M2-M5
               &$+$                      & \, $q=0:-$ \\
\hline
               &                      &\, $q=4: +$ \\
 M5-M5       & 0                    & $q=3: 0$ \\
               &                      &\, $q=1,2: -$ \\
\hline
\end{tabular}
\end{table}

\begin{table}[H]\centering
\caption{The nature of long-range interaction between two stacks of parallel or non-parallel $p\, \& \, p'$-branes in $D=10$}
\begin{tabular}{|c|c|c|}
\hline
$D=10$ & Parallel & Non-parallel \\
\hline
               &                      & $ q=p-2\, (p\ge 2):0$ \\
 D$p$-D$p$       & 0                    & \, $q=p-1\, (p \ge 1) :+$          \\
               &                      &\quad\, $0\leqslant\, q \,<p-2:-$\quad \\
\hline
               &                      & $q=p-3\,(p \ge 3):0$ \\
D$p$-D$(p-2)$
               & $+$                      &\quad\, $0\leqslant q<p-3:-$ \\
\hline
D$p$-D$(p-4)$ & 0    & $-$ \\
\hline
D$p$-D$(p-6)$ & $-$ & $-$ \\
\hline
           & $p=0:0$             &   \\
 D$p$-F1
           & \,$p \ge 1:+$             & 0   \\
\hline
                  & $p=6:0$      &    \\
D$p$-NS5 $(p>5)$
                  & \,$p>6:-$       & $-$     \\
\hline
                    &                  & $ q=p'-1\,(p' > 1) :0$ \\
D$p'$-NS5 $(p'<5)$
                    & $+$                  &\quad\,\, $0\leqslant q<p'-1:-$ \\
\hline
F1-NS5 & 0                  & $-$ \\
\hline
F1-F1 & 0                     & $+$ \\
\hline
           &                           & \,$q=2:-$ \\
 NS5-NS5 & 0             & $q=3:0$ \\
           &                           & $q=4:+$ \\
\hline
         &                             &\qquad  $q=4:0$ \\
D5-NS5
         &$+$                           & $2\leqslant q<4:-$ \\
\hline
                    & D1-D5:0 & D1-D5:$-$     \\
p\,$\&$\,(6-p) system & D0-D6:$-$ & D2-D4($q=0$):$-$ \\
                    & D2-D4:+ & D2-D4($q=1$):0 \\
                    & F-NS5:0 & F-NS5:$-$ \\
\hline
\end{tabular}
\end{table}

\begin{table}[H]\centering
\scriptsize
\caption{The nature of long-rang interaction between two stacks of non-parallel branes in $D=10$}
\begin{tabular}{|c|c|c|c|c|c|c|c|c|c|c|}
\hline
 & D1 & D2 & D3 & D4 & D5 & D6 & D7 & F1 & NS5  \\
\hline
D1 & $q=0:+$ & & & & & & & & \\
\hline
D2 & $\diagup$ & $q=0:0$ & & & & & & & \\
     &   & $q=1:+$ & & & & & & & \\
\hline
     &         &   & $q=0:-$ & & & & & & \\
D3 & $q=0:0$ & $\diagup$ & $q=1:0$ & & & & & & \\
     &         &   & $q=2:+$ & & & & & & \\
\hline
     &   &  &   & $q=0:-$ & & & & & \\
     &   & $q=1:0$ &   & $q=1:-$ & & & & & \\
D4 & $\diagup$ & $q=0:-$ & $\diagup$ & $q=2:0$            & & & & & \\
     &   &  &   & $q=3:+$            & & & & & \\
\hline
     &         &   & $q=0:-$            &   & $q=2:-$  & & & & \\\
D5 & $q=0:-$ & $\diagup$ &  $q=1:-$      & $\diagup$ & $q=3:0$  & & & & \\
     &         &   & $q=2:0$ &   & $q=4:+$  & & & & \\
\hline
     &   & $q=0:-$ &   & $q=2:-$ &   & $q=4:0$ & & &  \\
D6 &$\diagup$ &         &$\diagup$ &         & $\diagup$ &         & & &  \\
     &   & $q=1:-$ &   & $q=3:0$ &   & $q=5:+$ & & &  \\
\hline
D7 & $q=0:-$ & $\diagup$ & $q=2:-$ & $\diagup$ & $q=4:0$ & $\diagup$& $q=6:+$ & &  \\
\hline
F1 & $q=0:0$ & $q=0:0$ & $q=0:0$ & $q=0:0$ & $q=0:0$ & $q=0:0$ & $q=0:0$ & $q=0:+$ &  \\
\hline
      &         & $q=1:0$ & $q=0:-$ & $q=1:-$ & $q=2:-$ & $q=3:-$ &         &         & $q=2:-$  \\
NS5 & $q=0:0$ &         & $q=1:-$ & $q=2:-$ & $q=3:-$ &         & $q=4:-$ & $q=0:-$ & $q=3:0$  \\
      &         & $q=0:-$ & $q=2:0$ & $q=3:0$ & $q=4:0$ & $q=4:-$ &         &         & $q=4:+$  \\
\hline
\end{tabular}
\end{table}

\section{The long-range interactions between non-parallel D-branes with flux}\label{non-parallel D-brane with flux}

This section is an application of the results ontained in the previous sections. We here specify the discussion to D-branes. As we know that D-branes carrying a single abelian electric or magnetic world volume flux are the non-threshold BPS (F, D$p$) bound state \cite{Witten:1996np, Arfaei:1998np,Lu:1999jhep, Lu:1999np, Hashimoto:1997pr, DiVecchia:2000np} or the non-threshold BPS (D$(p-2)$, D$p$) bound state \cite{Breckenridge:1996pr, Costa:1996zd, DiVecchia:1997np}, respectively. The long-range interaction between two parallel D-branes with each carrying an abelian world-volume flux has been discussed in \cite{lu ning wei xu: interaction between}.
There are also other D-brane bound states (D$(p- 2k)$, D$p$) with $k = 2, 3, 4$, respectively. For each given $k$, the corresponding bound state can be obtained by turning on the non-abelian world volume magnetic fluxes on the Dp-branes such that $\int F^i \neq 0$ only for $i = k$ with $F^i = F\wedge \cdots \wedge F$, where the number of wedge products is $i -1$ \cite{Taylor:1997ay}.  The corresponding interaction between two such D-brane configurations  placed parallel has been discussed in \cite{Wu and Wang}. Like the $k=1$ case, these bound states correspond to the delocalized D$(p-4)$-branes within D$p$-branes, D$(p-6)$-branes within D$p$-branes and D0-branes within D8-branes. Therefore, the interaction between two stacks of such brane bound states, placed parallel or non-parallel, can be obtained as the sum of interactions between the constituent simple branes in the two bound states. For example, the interaction between (D$(p - 2)$, D$p$) and (D$(p - 6)$, D$(p - 4)$) ($p \ge 6$) can be obtained as the sum of interactions between the D$(p - 2)$ and D$(p - 6)$,  D$(p - 2)$ and D$(p -4)$, D$p$ and D$(p - 6)$, and D$p$ and D$(p - 4)$. Given what we obtain in the previous sections and the classification given in section 5 for the interaction of simple D-branes,  the interaction between two sets of constituent simple branes in the two bound states can be read simply from Table 3 given in section 5 and the nature of the interaction (attractive, repulsive or vanishing) between two such bound states can therefore be determined accordingly without the need of detail calculations.

As a simple illustration of the above, we will focus on the long-range interaction between two branes placed non-parallel with each carrying a $k=1$ constant magnetic flux, which corresponds to the non-threshold BPS (D$(p-2)$, D$p$) bound states. The total number for various possibilities is 155 and the results are given in the Appendix. For example, we consider the interaction between the bound state (D$4'$, D6) and the bound state (D$2'$, D4) with the constituent D6 branes non-parallel to D4 with the corresponding $q = 2, 3$. For each case, the interaction nature can be determined from the table 3 for simple constituent branes. All possible cases are listed in Table 4 and Table 5.

Concretely, we specify the D6-branes along $x^1,x^2,x^3,x^4,x^5,x^6$ and the D$4'$ in the bound state along $x^1,x^2,x^3,x^4$ but delocalized along $x^5, x^6$. In the following, we will fix the spatial directions for D6 and D$4'$ as specified above and let the spatial directions for the D$2'$ and D4 in the bound state (D$2'$, D4) vary. We discuss one specific case here with $q = 2$ as an example. Consider the D4 along $x^5, x^6, x^7, x^8$ and D$2'$ along $x^7, x^8$ for the bound state
(D$2'$, D4), i.e., the first case in Table 4. From Table 3, we know that the interaction between the constituent D4 and D6 is repulsive, the one between D4 and $D4'$ also repulsive, the one between D$2'$ and D$4'$ also repulsive, and finally the one between D$2'$ and D6 repulsive. So we must conclude that the total interaction between two such bound state is also repulsive.

Note that in both Table 4 and 5, we use ``${\rm x}$" to denote a spatial direction shared by the two constituent branes in a bound state  and ``$\tilde {\rm x}$" to denote the direction not shared by the constituent branes.   We use the question mark ``?" to denote the cases for which the total interaction cannot be determined by the prescription given above and we must do the computations to determine the nature of the underlying long-range interaction. This can be done without much difficulty following the couplings given in \cite{Wu and Wang, lu ning wei xu: interaction between}. There actually exists only one case for either $q = 2$ or $q = 3$ as shown in the corresponding table. We just quote the couplings needed for this computations from \cite{Wu and Wang, lu ning wei xu: interaction between} as
\begin{equation}\label{eq: effective couplings_D brane with flux}
  J_{h}^{(i)}=-c_p V_{p+1}\text{Tr}\left\{\sqrt{-det(\eta+\hat{F})}\left[(\eta+\hat{F})^{-1}\right]\right\}^{\alpha\beta}h_{\beta\alpha}
\end{equation}
for the graviton,
\begin{equation}
  J_{B}^{(i)}=-\frac{c_p V_{p+1}}{\sqrt{2}}\text{Tr}\left\{\sqrt{-det(\eta+\hat{F})}\left[(\eta+\hat{F})^{-1}\right]\right\}^{\alpha\beta}B_{\beta\alpha}
\end{equation}
for the Kalb-Ramond field,
\begin{equation}
  J_{\phi}^{(i)}=\frac{c_p V_{p+1}}{2\sqrt{2}} \text{Tr}\left\{\sqrt{-det(\eta+\hat{F})}\left[(3-p)\mathbf{1}_{N}+tr\left(\hat{F}(\eta+\hat{F})^{-1}\right)\right]\right\}\phi
\end{equation}
for the dilaton,
\begin{equation}
  J_{A_{p+1}}^{(i)}=\frac{\sqrt{2}c_p V_{p+1}N_i}{(p+1)!}A_{\alpha_{0}\alpha_{1}\cdots\alpha_{p}}\epsilon^{\alpha_{0}\alpha_{1}\cdots\alpha_{p}}
\end{equation}
for the R-R potential $A_{p+1}$, and
\begin{equation}
  J_{A_{p+1-2k}}^{(i)}=\frac{\sqrt{2}c_p V_{p+1}}{2^{k}k!(p+1-2k)!}\text{Tr}\{\hat{F}_{\alpha_{0}\alpha_{1}}
  \cdots\hat{F}_{\alpha_{2k-2}\alpha_{2k-1}}\}A_{\alpha_{2k}\alpha_{2k+1}\cdots\alpha_{p}}\epsilon^{\alpha_{0}\alpha_{1}\cdots\alpha_{p}}
\end{equation}
for the R-R potential $A_{p+1-2k}$.

\begin{table}[H]\centering
\caption{Interaction between (D$2'$,D4) and (D$4'$, D6)-branes with $q=2$}
\begin{tabular}{|c|c|c|c|c|c|c|c|c|c|l|c|}
\hline
  $q=2$&  $x^1$ & $x^2$ & $x^3$ & $x^4$ & $x^5$   & $x^6$   & $x^7$ & $x^8$ & $x^9$ &             &  \\
\hline
D6   &  x & x & x & x & \gx & \gx &   &   &   &             &  \\
D$4'$  &  x & x & x & x &     &     &   &   &   &             &  \\
\hline
\hline
D4   &    &   &   &   & \gx & \gx & x & x &   & D4-D6:$-$   & \multirow{4}*{$-$}  \\
       &       &   &   &   &     &     &   &   &   & D4-D$4'$:$-$  &    \\
D$2'$  &    &   &   &   &     &     & x & x &   & D$2'$-D6:$-$  &    \\
       &      &   &   &   &     &     &   &   &   & D$2'$-D$4'$:$-$ &    \\
\hline
D4   &    &   &   &   & \gx & x   & \gx & x &   & D4-D6:$-$   & \multirow{4}*{$-$}  \\
       &      &   &   &   &     &     &     &   &   & D4-D$4'$:$-$  &    \\
D$2'$  &    &   &   &   &     & x   &     & x &   & D$2'$-D6:$-$  &    \\
       &      &   &   &   &     &     &     &   &   & D$2'$-D$4'$:$-$ &    \\
\hline
D4   & \gx&   &   &   & \gx &     & x   & x &   & D4-D6:$-$   & \multirow{4}*{$-$}  \\
       &      &   &   &   &     &     &     &   &   & D4-D$4'$:$-$  &    \\
D$2'$  &    &   &   &   &     &     & x   & x &   & D$2'$-D6:$-$  &    \\
       &      &   &   &   &     &     &     &   &   & D$2'$-D$4'$:$-$ &    \\
\hline
D4  &  x &   &   &   & \gx &     & \gx & x &   & D4-D6:$-$   & \multirow{4}*{$-$}  \\
       &      &   &   &   &     &     &     &   &   & D4-D$4'$:$-$  &    \\
D$2'$  &  x &   &   &   &     &     &     & x &   & D$2'$-D6:$-$  &    \\
       &      &   &   &   &     &     &     &   &   & D$2'$-D$4'$:0 &    \\
\hline
D4   &    &   &   &   & x   &  x  & \gx &\gx&   & D4-D6:$-$   & \multirow{4}*{$-$}  \\
       &      &   &   &   &     &     &     &   &   & D4-D$4'$:$-$  &    \\
D$2'$  &    &   &   &   & x   &  x  &     &   &   & D$2'$-D6:0  &    \\
       &      &   &   &   &     &     &     &   &   & D$2'$-D$4'$:$-$ &    \\
\hline
D4   & \gx&   &   &   & x   &     & \gx & x &   & D4-D6:$-$   & \multirow{4}*{$-$}  \\
       &      &   &   &   &     &     &     &   &   & D4-D$4'$:$-$  &    \\
D$2'$  &    &   &   &   & x   &     &     & x &   & D$2'$-D6:$-$  &    \\
       &      &   &   &   &     &     &     &   &   & D$2'$-D$4'$:$-$ &    \\
\hline
D4   &  x &   &   &   & x   &     & \gx &\gx&   & D4-D6:$-$   & \multirow{4}*{$-$}  \\
       &      &   &   &   &     &     &     &   &   & D4-D$4'$:$-$  &    \\
D$2'$  &  x &   &   &   & x   &     &     &   &   & D$2'$-D6:0  &    \\
       &     &   &   &   &     &     &     &   &   & D$2'$-D$4'$:0 &    \\
\hline
D4   & \gx&\gx&   &   &     &     &  x  & x &   & D4-D6:$-$   & \multirow{4}*{$-$}  \\
       &      &   &   &   &     &     &     &   &   & D4-D$4'$:0  &    \\
D$2'$  &    &   &   &   &     &     &  x  & x &   & D$2'$-D6:$-$  &    \\
       &      &   &   &   &     &     &     &   &   & D$2'$-D$4'$:$-$ &    \\
\hline
D4   & \gx& x &   &   &     &     &\gx  & x &   & D4-D6:$-$   & \multirow{4}*{$-$}  \\
       &      &   &   &   &     &     &     &   &   & D4-D$4'$:0  &    \\
D$2'$  &    & x &   &   &     &     &     & x &   & D$2'$-D6:$-$  &    \\
       &      &   &   &   &     &     &     &   &   & D$2'$-D$4'$:0 &    \\
\hline
D4   &  x & x &   &   &     &     &\gx  &\gx&   & D4-D6:$-$   & \multirow{4}*{$?$}  \\
       &      &   &   &   &     &     &     &   &   & D4-D$4'$:0  &    \\
D$2'$  &  x & x &   &   &     &     &     &   &   & D$2'$-D6:0  &    \\
       &      &   &   &   &     &     &     &   &   & D$2'$-D$4'$:+ &    \\
\hline
\end{tabular}
\end{table}

\begin{table}[H]\centering
\caption{Interaction between (D$2'$, D4) and (D$4'$, D6)-branes with $q=3$}
\begin{tabular}{|c|c|c|c|c|c|c|c|c|c|l|c|}
\hline
  $q=3$&  $x^1$ & $x^2$ & $x^3$ & $x^4$ & $x^5$   & $x^6$   & $x^7$ & $x^8$ & $x^9$ &             &  \\
\hline
D6   &  x & x & x & x & \gx & \gx &   &   &   &             &  \\
D$4'$  &  x & x & x & x &     &     &   &   &   &             &  \\
\hline
\hline
D4   &  x &   &   &   & \gx & \gx & x &   &   & D4-D6:0   & \multirow{4}*{$-$}  \\
       &      &   &   &   &     &     &   &   &   & D4-D$4'$:$-$  &    \\
D$2'$  &  x &   &   &   &     &     & x &   &   & D$2'$-D6:$-$  &    \\
       &      &   &   &   &     &     &   &   &   & D$2'$-D$4'$:0 &    \\
\hline
D4   & \gx&   &   &   & \gx & x   &  x  &   &   & D4-D6:0   & \multirow{4}*{$-$}  \\
       &      &   &   &   &     &     &     &   &   & D4-D$4'$:$-$  &    \\
D$2'$  &    &   &   &   &     & x   &  x  &   &   & D$2'$-D6:$-$  &    \\
       &      &   &   &   &     &     &     &   &   & D$2'$-D$4'$:$-$ &    \\
\hline
D4   &  x &   &   &   & \gx &  x  &\gx  &   &   & D4-D6:0   & \multirow{4}*{$-$}  \\
       &      &   &   &   &     &     &     &   &   & D4-D$4'$:$-$  &    \\
D$2'$  &  x &   &   &   &     &  x  &     &   &   & D$2'$-D6:0  &    \\
       &      &   &   &   &     &     &     &   &   & D$2'$-D$4'$:0 &    \\
\hline
D4   &  x &\gx&   &   & \gx &     &  x  &   &   & D4-D6:0   & \multirow{4}*{$-$}  \\
       &      &   &   &   &     &     &     &   &   & D4-D$4'$:0  &    \\
D$2'$  &  x &   &   &   &     &     &  x  &   &   & D$2'$-D6:$-$  &    \\
       &      &   &   &   &     &     &     &   &   & D$2'$-D$4'$:0 &    \\
\hline
D4   &  x & x &   &   &\gx  &     & \gx &   &   & D4-D6:0   & \multirow{4}*{$+$}  \\
       &      &   &   &   &     &     &     &   &   & D4-D$4'$:0  &    \\
D$2'$  &  x & x &   &   &     &     &     &   &   & D$2'$-D6:0  &    \\
       &      &   &   &   &     &     &     &   &   & D$2'$-D$4'$:+ &    \\
\hline
D4   & \gx&\gx&   &   & x   &     &  x  &   &   & D4-D6:0   & \multirow{4}*{$-$}  \\
       &      &   &   &   &     &     &     &   &   & D4-D$4'$:0  &    \\
D$2'$  &    &   &   &   & x   &     &  x  &   &   & D$2'$-D6:$-$  &    \\
       &      &   &   &   &     &     &     &   &   & D$2'$-D$4'$:$-$ &    \\
\hline
D4   & \gx& x &   &   & x   &     & \gx &   &   & D4-D6:0   & \multirow{4}*{$0$}  \\
       &     &   &   &   &     &     &     &   &   & D4-D$4'$:0  &    \\
D$2'$  &    & x &   &   & x   &     &     &   &   & D$2'$-D6:0  &    \\
       &      &   &   &   &     &     &     &   &   & D$2'$-D$4'$:0 &    \\
\hline
D4   & \gx&   &   &   & x   &  x  & \gx &   &   & D4-D6:0   & \multirow{4}*{$-$}  \\
       &      &   &   &   &     &     &     &   &   & D4-D$4'$:$-$  &    \\
D$2'$  &    &   &   &   & x   &  x  &     &   &   & D$2'$-D6:0  &    \\
       &      &   &   &   &     &     &     &   &   & D$2'$-D$4'$:$-$ &    \\
\hline
D4   & \gx&\gx& x &   &     &     & x   &   &   & D4-D6:0   & \multirow{4}*{$?$}  \\
       &      &   &   &   &     &     &     &   &   & D4-D$4'$:+  &    \\
D$2'$  &    &   & x &   &     &     & x   &   &   & D$2'$-D6:$-$  &    \\
       &      &   &   &   &     &     &     &   &   & D$2'$-D$4'$:0 &    \\
\hline
D4   & \gx& x & x &   &     &     &\gx  &   &   & D4-D6:0   & \multirow{4}*{$+$}  \\
       &      &   &   &   &     &     &     &   &   & D4-D$4'$:+  &    \\
D$2'$  &    & x & x &   &     &     &     &   &   & D$2'$-D6:0  &    \\
       &      &   &   &   &     &     &     &   &   & D$2'$-D$4'$:+ &    \\
\hline
\end{tabular}
\end{table}

\newpage

\noindent
 In the above, each of these bulk fields is a singlet under the $U(N)$ gauge group, $\hat F=2\pi\alpha' F$ is the constant world-volume flux and is an $N\times N$ matrix under the $U(N)$. ``$\text{Tr}$" is with respect to the trace in $N\times N$ space, and $c_{p}=T_{p}\kappa /g_{s}=T_{p}\kappa_{10}$. $\mathbf{1}_{N}$ stands for the $N \times N$ unit matrix while $0_N$ stands for the $N\times N$ zero-matrix as will be used in the following representation for $\hat F$.

Consider the $q=2$ case marked with $``?"$ mark in Table 4. The constant fluxes $\hat F_1$ and $\hat F_2$ on the world-volume of D6- and D4-branes are chosen accordingly, respectively, as:
\begin{align}
\hat F_1=&
\begin{pmatrix}
  0_{N_1} &    &    &    &    &    & \\
     & 0_{N_1} &    &    &    &    &  \\
     &    & 0_{N_1} &    &    &    &  \\
     &    &    & 0_{N_1} &    &    &  \\
     &    &    &    & 0_{N_1} &    &  \\
     &    &    &    &    & 0_{N_1} & - g_1\mathbf1_{N_1} \\
     &    &    &    &    & g_1\mathbf1_{N_1} & 0_{N_1} \\
\end{pmatrix}\nonumber\\
\hat F_2=&
\begin{pmatrix}
  0_{N_2} &    &    &    &    &    &    &    &   \\
     & 0_{N_2} &    &    &    &    &    &    &    \\
     &    & 0_{N_2} &    &    &    &    &    &    \\
     &    &    & \diagup &    &    &    &    &    \\
     &    &    &    & \diagup &    &    &    &    \\
     &    &    &    &    & \diagup &    &    &   \\
     &    &    &    &    &  & \diagup &   &   \\
     &    &    &    &    &    &    & 0_{N_2} &  -g_2\mathbf1_{N_2}  \\
     &    &    &    &    &    &    &  g_2\mathbf1_{N_2}  & 0_{N_2} \\
\end{pmatrix}
\end{align}
The ``$\diagup$" in the matrix of $\hat F_2$ means nothing should be here and should not be counted in the computations, because the D4-branes are not along those directions.  The explicit couplings for the two bound states are:
\begin{align}
  J_h^{(1)}&=-c_6V_7N_1\sqrt{1+g_1^2}V_1^{\alpha\beta}h_{\alpha\beta}\nonumber\\
  J_\phi^{(1)}&=-\frac{\sqrt2}{4}c_6V_7N_1\sqrt{1+g_1^2}\left( 3-\frac{2g_1^2}{1+g_1^2} \right)\phi
\end{align}
for D6-branes carrying the flux $\hat F_1$, and
\begin{align}
  J_h^{(2)}&=-c_4V_5N_2\sqrt{1+g_2^2}V_2^{\alpha\beta}h_{\alpha\beta}\nonumber\\
  J_\phi^{(2)}&=-\frac{\sqrt2}{4}c_4V_5N_2\sqrt{1+g_2^2}\left( 1-\frac{2g_2^2}{1+g_2^2} \right)\phi
\end{align}
for D4-branes carrying the flux $\hat F_2$. Where $V_1$ and $V_2$ are
\begin{align}
V_1^{\alpha\beta}&=
\begin{pmatrix}
  -1 &    &    &    &    &    &   \\
     & 1 &    &    &    &    &    \\
     &    & 1 &    &    &    &    \\
     &    &    & 1 &    &    &    \\
     &    &    &    & 1 &    &    \\
     &    &    &    &    & \frac1{1+g_1^2} & \frac{g_1}{1+g_1^2} \\
     &    &    &    &    & \frac{-g_1}{1+g_1^2} & \frac1{1+g_1^2}\\
\end{pmatrix}\nonumber\\
V_2^{\alpha\beta}&=
\begin{pmatrix}
  -1 &    &    &    &    &    &    &    &   \\
     & 1 &    &    &    &    &    &    &    \\
     &    & 1 &    &    &    &    &    &    \\
     &    &    & \diagup &    &    &    &    &    \\
     &    &    &    & \diagup &    &    &    &    \\
     &    &    &    &    & \diagup &    &    &   \\
     &    &    &    &    &  & \diagup &   &   \\
     &    &    &    &    &    &    & \frac1{1+g_2^2} &  \frac{g_2}{1+g_2^2}  \\
     &    &    &    &    &    &    &  \frac{-g_2}{1+g_2^2}  & \frac{1}{1+g_2^2} \\
\end{pmatrix}
\end{align}
 Just like the case in the absence of fluxes, we don't have the contributions from the R-R field exchanges for the non-parallel case, i.e., $U_{A_{p+1}}=0$ and $U_{A_{p-1}}=0$. Further since the fluxes on D6- and D4-branes are along different directions in the ``?" case considered, so we have $U_B =0$.

Therefore the long-range interactions due to the exchanges of graviton and dilaton, respectively, are
\begin{align}
  U_h&=\frac1{8k_\perp^2}c_6c_4V_7V_5N_1N_2\sqrt{1+g_1^2}\sqrt{1+g_2^2}\left[ 24-\left(5+\frac2{1+g_1^2}\right)\left(3+\frac2{1+g_2^2}\right) \right], \\
  U_\phi&=\frac1{8k_\perp^2}c_6c_4V_7V_5N_1N_2\sqrt{1+g_1^2}\sqrt{1+g_2^2}\left( 3-\frac{2g_1^2}{1+g_1^2} \right)\left( 1-\frac{2g_2^2}{1+g_2^2} \right).
\end{align}
Then the total interaction is
\begin{align}
  U=U_{h}+U_{\phi}=\frac1{k_\perp^2}c_6c_4V_7V_5N_1N_2\sqrt{1+g_1^2}\sqrt{1+g_2^2}\frac{g_1^2g_2^2-1}{(1+g_1^2)(1+g_2^2)}.
\end{align}
Therefore for the marked ``?" case in $q=2$, if $|g_1g_2| >1$, the interaction between the two bound states is attractive, if $|g_1g_2| =1$, there is no interaction, and if $|g_1g_2| <1$, the interaction is repulsive.

For the marked ``?" case in $q=3$  given in Table 5, the total interaction can also be obtained similarly  and the result is
\begin{align}
  U=U_{h}+U_{\phi}=\frac1{k_\perp^2}c_6c_4V_7V_5N_1N_2\sqrt{1+g_1^2}\sqrt{1+g_2^2}\frac{g_1^2-g_2^2}{(1+g_1^2)(1+g_2^2)},
\end{align}
and therefore the nature of the interaction depends on the relative magnitude of $|g_1|$ and $|g_2|$.

\section{Summary}\label{summary}

In this paper, we calculate the interaction between two stacks of  non-parallel simple branes of the same or different type at a separation for various cases in $D=10$ and $D=11$ via the effective field theory approach \cite{Wu and Wang, Prof. Lu's lecture notes}. In some special cases such as those involved p-branes and their Hodge-dual objects, the explicit results can also be given in diverse dimensions. Combined with the known parallel cases \cite{Wu and Wang, Prof. Lu's lecture notes, lu ning wei xu: interaction between}, we give a classification of long-range interactions for simple branes. This serves as a dictionary and is useful when we consider more complicated brane systems such as the brane bound states for which the nature of interaction (attractive or repulsive) can often be simply read from this dictionary without the need of detail computations.

We demonstrate this explicitly using an example for a particular case for the interaction between the (D$4'$, D6) and the (D$2'$, D4). For the two systems, there are ten cases for either $q = 2$ or $q = 3$ but there is only one case for either of them for which the interaction cannot be simply read from the dictionary. We then calculate the corresponding interaction using the couplings derived in \cite{Wu and Wang,lu ning wei xu: interaction between} and discuss how the nature of interaction depends on the fluxes on the branes. We also provide two tables given in the appendix showing the nature of the long-range interaction between two stacks of D-branes placed non-parallel with each carrying a constant $k=1$ magnetic flux.

\section*{\noindent{Acknowledgements}}

We would like to thank J. X. Lu, for suggesting us this work and help us improve this manuscript. We would also like to thank Yi Sun and Wei Gu for helpful discussions. This work is supported by  a key
grant from the NSF of China with Grant No : 11235010.

\section*{\centering{Appendix}}

In this Appendix, we list all the possibilities showing the nature of the long-range interaction between two D-branes placed non-parallel with each carrying a constant $k=1$ magnetic flux. For this purpose, each D$p$-brane must have at least two spatial world-volume directions, i.e., $p\ge 2$. Further, as mentioned earlier, to avoid possible complication and a reasonable good behavior of
the interaction, we limit ourselves in the following to the $p \le 7$ cases.   We use ``${\rm x}$" here to denote the spatial direction shared by the two constituent branes in a bound state  and ``$\tilde {\rm x}$" to denote the direction not shared by the constituent branes (``$\tilde {\rm x}$" also means the direction that the magnetic flux occupies). In addition, we use the signs $``+"$, $``-"$ and ``0" to to denote the attractive, repulsive and vanishing interactions, respectively.  Here again ``?" indicates a case for which the total interaction cannot be simply determined by the prescription given earlier and explicit computations are needed to determine the nature of the underlying long-range interaction. Note that $x^9$ is the spatial direction transverse to both stacks of branes and doesn't appear in the following table.
\newpage

\begin{longtable}{|c|c|c|c|c|c|c|c|c||c|c|c|c|c|c|c|c|c||c|c|}
\caption{The long-range interaction between two non-parallel D-branes in Type IIB theory with $k=1$ magnetic flux}
\\ \hline
\multirow{2}*{D$p$} & \multicolumn{8}{c|}{location and flux}& \multirow{2}*{D$p'$} & \multicolumn{8}{c|}{location and flux} &\multirow{2}*{$q$} & \multirow{2}*{$U$}\\ \cline{2-9}\cline{11-18}
  & 1 & 2 & 3 & 4 & 5 & 6 & 7 & 8 & & 1 & 2 & 3 & 4 & 5 & 6 & 7 & 8 & & \\ \hline
\endfirsthead
\hline
\multirow{2}*{D$p$} & \multicolumn{8}{c|}{location and flux}& \multirow{2}*{D$p'$} & \multicolumn{8}{c|}{location and flux} &\multirow{2}*{$q$} & \multirow{2}*{$U$}\\ \cline{2-9}\cline{11-18}
  & 1 & 2 & 3 & 4 & 5 & 6 & 7 & 8 & & 1 & 2 & 3 & 4 & 5 & 6 & 7 & 8 & & \\ \hline \endhead
\hline
\multicolumn{20}{r}{\footnotesize{to be continued on the next page}} \\ \endfoot
\hline
\endlastfoot
D7 & x & x & x & x & x &\gx&\gx&   &   D7            & x & x & x & x &\gx&\gx&   & x &  $q=6$ & + \\ \cline{11-18}\cline{20-20}
   &   &   &   &   &   &   &   &   &                 & x & x & x & x & x &\gx&   &\gx& & + \\ \cline{11-18}\cline{20-20}
   &   &   &   &   &   &   &   &   &                 & x & x & x &\gx&\gx& x &   & x & & + \\ \cline{11-18}\cline{20-20}
   &   &   &   &   &   &   &   &   &                 & x & x & x & x &\gx& x &   &\gx& &+ \\ \cline{11-18}\cline{20-20}\cline{2-9}
   & x & x & x & x &\gx&\gx& x &   &                 & x & x & x & x &\gx&\gx&   & x & &+ \\ \cline{11-18}\cline{20-20}
   &   &   &   &   &   &   &   &   &                 & x & x & x &\gx& x &\gx&   & x & &+ \\ \cline{11-18}\cline{20-20}
   &   &   &   &   &   &   &   &   &                 & x & x & x & x & x &\gx&   &\gx& &+ \\ \cline{11-18}\cline{20-20}
   &   &   &   &   &   &   &   &   &                 & x & x &\gx&\gx& x & x &   & x & &? \\ \cline{11-18}\cline{20-20}
   &   &   &   &   &   &   &   &   &                 & x & x & x &\gx& x & x &   &\gx& &+ \\ \hline\hline
D7 & x & x & x & x &\gx&\gx& x &   &   D5            & x & x &\gx&\gx&   &   &   & x & $q=4$ &? \\ \cline{11-18}\cline{20-20}
   &   &   &   &   &   &   &   &   &                 & x & x & x &\gx&   &   &   &\gx& &+ \\ \cline{11-18}\cline{20-20}\cline{2-9}
   & x & x & x &\gx&\gx& x & x &   &                 & x & x &\gx&\gx&   &   &   & x & &$-$ \\ \cline{11-18}\cline{20-20}
   &   &   &   &   &   &   &   &   &                 & x & x & x &\gx&   &   &   &\gx& &+ \\ \cline{11-18}\cline{20-20}
   &   &   &   &   &   &   &   &   &                 & x &\gx&\gx& x &   &   &   &\gx& &$-$ \\ \cline{11-18}\cline{20-20}
   &   &   &   &   &   &   &   &   &                 & x & x &\gx& x &   &   &   &\gx& &0 \\ \cline{11-18}\cline{20-20}\cline{2-9}
   & x & x &\gx&\gx& x & x & x &   &                 & x & x &\gx&\gx&   &   &   & x & &$-$ \\ \cline{11-18}\cline{20-20}
   &   &   &   &   &   &   &   &   &                 & x &\gx& x &\gx&   &   &   & x & &$-$ \\ \cline{11-18}\cline{20-20}
   &   &   &   &   &   &   &   &   &                 & x & x & x &\gx&   &   &   &\gx& &$-$ \\ \cline{11-18}\cline{20-20}
   &   &   &   &   &   &   &   &   &                 &\gx&\gx& x & x &   &   &   & x & &$-$ \\ \cline{11-18}\cline{20-20}
   &   &   &   &   &   &   &   &   &                 & x &\gx& x & x &   &   &   &\gx& &$-$ \\ \hline\hline
D7 & x & x &\gx&\gx& x & x & x &   &   D3            &\gx&\gx&   &   &   &   &   & x & $q=2$ &$-$ \\ \cline{11-18}\cline{20-20}
   &   &   &   &   &   &   &   &   &                 & x &\gx&   &   &   &   &   &\gx& &$-$ \\ \cline{11-18}\cline{20-20}\cline{2-9}
   & x &\gx&\gx& x & x & x & x &   &                 &\gx&\gx&   &   &   &   &   & x & &$-$ \\ \cline{11-18}\cline{20-20}
   &   &   &   &   &   &   &   &   &                 & x &\gx&   &   &   &   &   &\gx& &$-$ \\ \cline{11-18}\cline{20-20}
   &   &   &   &   &   &   &   &   &                 &\gx& x &   &   &   &   &   &\gx& &$-$ \\ \cline{11-18}\cline{20-20}\cline{2-9}
   &\gx&\gx& x & x & x & x & x &   &                 &\gx&\gx&   &   &   &   &   & x & &$-$ \\ \cline{11-18}\cline{20-20}
   &   &   &   &   &   &   &   &   &                 & x &\gx&   &   &   &   &   &\gx& &$-$ \\ \hline\hline
D5 & x & x & x &\gx&\gx&   &   &   &   D5            & x & x &\gx&\gx&   & x &   &   &$q=4$ &+ \\ \cline{11-18}\cline{20-20}
   &   &   &   &   &   &   &   &   &                 & x & x & x &\gx&   &\gx&   &   & &+ \\ \cline{11-18}\cline{20-20}
   &   &   &   &   &   &   &   &   &                 & x &\gx&\gx& x &   & x &   &   & &+ \\ \cline{11-18}\cline{20-20}
   &   &   &   &   &   &   &   &   &                 & x & x &\gx& x &   &\gx&   &   & &+ \\ \cline{11-18}\cline{20-20}\cline{2-9}
   & x & x &\gx&\gx& x &   &   &   &                 & x & x &\gx&\gx&   & x &   &   & &+ \\ \cline{11-18}\cline{20-20}
   &   &   &   &   &   &   &   &   &                 & x & \gx & \gx & x &     &  x   &      &       & &+    \\ \cline{11-18}\cline{20-20}
   &   &   &   &   &   &   &   &   &                 & x & x & \gx & x &     & \gx  &      &       &  &+   \\ \cline{11-18}\cline{20-20}
   &   &   &   &   &   &   &   &   &                 & \gx & \gx & x & x &     &  x   &      &       &  &?   \\ \cline{11-18}\cline{20-20}
   &   &   &   &   &   &   &   &   &                 & x & \gx & x & x &     &  \gx  &      &     &  &+ \\ \cline{11-20}\cline{2-9}
   & x & x & x &\gx&\gx&   &   &   &                 & x & \gx & \gx &   &     & x  & x  &       & $q=3$&?    \\ \cline{11-18}\cline{20-20}
   &   &   &   &   &   &   &   &   &                 & x & x & \gx &   &     & \gx  & x  &       &  &+   \\ \cline{11-18}\cline{20-20}
   &   &   &   &   &   &   &   &   &                 & x & x & x &   &  & \gx  & \gx &       &  &+   \\ \cline{11-18}\cline{20-20}\cline{2-9}
   & x & x &\gx&\gx& x &   &   &   &                 & x & \gx& \gx &  &     &  x  &  x    &       & &$-$    \\ 
D5 &   &   &   &   &   &   &   &   &   D5            & x  & x  & \gx &   &     & \gx  & x  &     &$q=3$ &+    \\ \cline{11-18}\cline{20-20}
   &   &   &   &   &   &   &   &   &                 & \gx & \gx & x &   &     &  x  & x  &       &  &$-$   \\ \cline{11-18}\cline{20-20}
   &   &   &   &   &   &   &   &   &                 & x & \gx & x &   &     &  \gx  & x  &       & &0    \\ \cline{11-18}\cline{20-20}
   &   &   &   &   &   &   &   &   &                 & x & x & x &   &     &  \gx  & \gx &    &  &+   \\ \cline{11-18}\cline{20-20}\cline{2-9}
   & x &\gx&\gx& x & x &   &   &   &                 & \gx &\gx& x &   &     &  x  & x  &       &  &$-$   \\ \cline{11-18}\cline{20-20}
   &   &   &   &   &   &   &   &   &                 & x & \gx& x &   &     & \gx  & x  &       &  &$-$   \\ \cline{11-18}\cline{20-20}
   &   &   &   &   &   &   &   &   &                 & \gx & x & x &   &     & \gx  & x  &       &  &$-$   \\ \cline{11-18}\cline{20-20}
   &   &   &   &   &   &   &   &   &                 & x & x & x &   &     & \gx  & \gx &       &  &?   \\ \cline{11-18}\cline{20-20}
   &   &   &   &   &   &   &   &   &                 & x & \gx & \gx &   &    & x  &  x  &       & & $-$ \\ \cline{11-20}\cline{2-9}
   & x & x &\gx&\gx& x &   &   &   &                 & \gx & \gx &   &   &     &  x  &  x & x  & $q=2$ &$-$   \\ \cline{11-18}\cline{20-20}
   &   &   &   &   &   &   &   &   &                 & x & \gx &   &   &     & \gx &  x & x  &  &$-$   \\ \cline{11-18}\cline{20-20}
   &   &   &   &   &   &   &   &   &                 & x & x &   &   &     & \gx & \gx & x & &? \\ \cline{11-18}\cline{20-20}\cline{2-9}
   & x &\gx&\gx& x & x &   &   &   &                 &\gx&\gx&   &   &    &  x  &  x  &  x   &  &$-$   \\ \cline{11-18}\cline{20-20}
   &   &   &   &   &   &   &   &   &                 & x & \gx &   &   &     &  \gx & x  & x  &  &$-$   \\ \cline{11-18}\cline{20-20}
   &   &   &   &   &   &   &   &   &                 & \gx & x  &   &   &     & \gx& x  &  x  &  &$-$   \\ \cline{11-18}\cline{20-20}
   &   &   &   &   &   &   &   &   &                 & x & x &   &   &     & \gx & \gx & x  &  &$-$   \\ \cline{11-18}\cline{20-20}\cline{2-9}
   &\gx&\gx& x & x & x &   &   &   &                 & \gx & \gx &   &   &     &  x  & x  & x  & &$-$    \\ \cline{11-18}\cline{20-20}
   &   &   &   &   &   &   &   &   &                 & x & \gx &   &   &     & \gx & x  &  x  &  &$-$   \\ \cline{11-18}\cline{20-20}
   &   &   &   &   &   &   &   &   &                 & x & x &   &   &    & \gx & \gx & x  &  &$-$   \\ \hline\hline
D5 & x & x &\gx&\gx& x &   &   &   &   D3            & \gx & \gx &   &   &     &  x  &      &       & $q=2$ &?   \\ \cline{11-18}\cline{20-20}
   &   &   &   &   &   &   &   &   &                 & x & \gx &   &   &     &  \gx  &    &   & & +   \\ \cline{11-18}\cline{20-20}\cline{2-9}
   & x &\gx&\gx& x & x &   &   &   &                 & \gx & \gx &   &   &     &  x  &      &       & & $-$   \\ \cline{11-18}\cline{20-20}
   &   &   &   &   &   &   &   &   &                 & x & \gx  &   &   &    & \gx  &      &       & & +   \\ \cline{11-18}\cline{20-20}
   &   &   &   &   &   &   &   &   &                 & \gx  & x  &   &   &   & \gx &      &   &  &0   \\ \cline{11-18}\cline{20-20}\cline{2-9}
   &\gx&\gx& x & x & x &   &   &   &                 & \gx  & \gx &   &   &   & x  &      &       &  &$-$   \\ \cline{11-18}\cline{20-20}
   &   &   &   &   &   &   &   &   &                 & x & \gx &   &   &   & \gx &      &       &  &$-$   \\ \cline{11-20}\cline{2-9}
   & x &\gx&\gx& x & x &   &   &   &                 & \gx &   &   &   &   & \gx &  x  &       & $q=1$ &$-$   \\ \cline{11-18}\cline{20-20}
   &   &   &   &   &   &   &   &   &                 & x  &   &   &   &   &  \gx  &  \gx  &   & &?    \\ \cline{11-18}\cline{20-20}\cline{2-9}
   &\gx&\gx& x & x & x &   &   &   &                 &  \gx&   &   &   &     &   \gx   &   x   &   & &$-$    \\ \cline{11-18}\cline{20-20}
   &   &   &   &   &   &   &   &   &                 &  x &   &   &   &     &   \gx &   \gx  &     & & $-$ \\ \cline{11-20}\cline{2-9}
   &\gx&\gx& x & x & x &   &   &   &                 &   &   &   &   &     & \gx  & \gx &  x &$q=0$ &$-$    \\ \hline\hline
D3 & x &\gx&\gx&   &   &   &   &   &   D3            &\gx&\gx &   & x  &     &      &      &     &$q=2$ & +   \\ \cline{11-18}\cline{20-20}
   &   &   &   &   &   &   &   &   &                 & x  &\gx&   & \gx&     &      &      &     & & +   \\ \cline{11-18}\cline{20-20}
   &   &   &   &   &   &   &   &   &                 & \gx & x  &   &\gx &     &      &   &   & & +   \\ \cline{11-18}\cline{20-20}\cline{2-9}
   &\gx&\gx& x &   &   &   &   &   &                 &\gx &\gx &   & x  &     &      &      &     &  &+   \\ \cline{11-18}\cline{20-20}
   &   &   &   &   &   &   &   &   &                 & x &\gx &   & \gx &     &      &      &     &  &+   \\ \cline{11-20}\cline{2-9}
   & x &\gx&\gx&   &   &   &   &   &                 &\gx &  &   &\gx & x  &      &      &       & $q=1$ &+   \\ \cline{11-18}\cline{20-20}
   &   &   &   &   &   &   &   &   &                 & x &   &   &\gx  &\gx &      &      &   & &+    \\ \cline{11-18}\cline{20-20}\cline{2-9}
   &\gx&\gx& x &   &   &   &   &   &                 &\gx &   &   &\gx &  x &      &      &       &  &+   \\ \cline{11-18}\cline{20-20}
   &   &   &   &   &   &   &   &   &                 & x &   &   &\gx &\gx &      &      &       &  &+   \\ \cline{11-20}\cline{2-9}
   &\gx&\gx& x &   &   &   &   &   &                &   &   &   & \gx & \gx & x &      &       &  $q=0$ &?  \\ \hline
\end{longtable}

\newpage

\begin{longtable}{|c|c|c|c|c|c|c|c|c||c|c|c|c|c|c|c|c|c||c|c|}
\caption{The long-range interaction between two non-parallel D-branes in Type IIA theory with $k=1$ magnetic flux}
\\ \hline
\multirow{2}*{D$p$} & \multicolumn{8}{c|}{location and flux}& \multirow{2}*{D$p'$} & \multicolumn{8}{c|}{location and flux} &\multirow{2}*{$q$} & \multirow{2}*{$U$}\\ \cline{2-9}\cline{11-18}
  & 1 & 2 & 3 & 4 & 5 & 6 & 7 & 8 & & 1 & 2 & 3 & 4 & 5 & 6 & 7 & 8 & & \\ \hline
\endfirsthead
\hline
\multirow{2}*{D$p$} & \multicolumn{8}{c|}{location and flux}& \multirow{2}*{D$p'$} & \multicolumn{8}{c|}{location and flux} &\multirow{2}*{$q$} & \multirow{2}*{$U$}\\ \cline{2-9}\cline{11-18}
  & 1 & 2 & 3 & 4 & 5 & 6 & 7 & 8 & & 1 & 2 & 3 & 4 & 5 & 6 & 7 & 8 & & \\ \hline \endhead
\hline
\multicolumn{20}{r}{\footnotesize{to be continued on the next page}} \\ \endfoot
\hline
\endlastfoot
D6 & x  & x  & x & x  &\gx &\gx &    &     &    D6       & x & x &\gx&\gx&  x  &   &  x &    & $q=5$ &+   \\ \cline{11-18}\cline{20-20}
   &    &    &   &    &    &    &    &     &             & x & x & x &\gx&\gx &    &  x  &   &  &+   \\ \cline{11-18}\cline{20-20}
   &    &    &   &    &    &    &    &     &             & x & x & x & x  &\gx &   & \gx  &   & &+   \\ \cline{11-18}\cline{20-20}
   &    &    &   &    &    &    &    &     &             & x & x & x & \gx& x  &    & \gx  &  & &+   \\ \cline{11-18}\cline{20-20}\cline{2-9}
   &  x & x  &x  &\gx &\gx & x  &    &     &             & x & x & x  &\gx&\gx &    &  x   &   & &+   \\ \cline{11-18}\cline{20-20}
   &    &    &   &    &    &    &    &     &             & x & x &\gx& x & \gx &   &  x   &    & &+   \\ \cline{11-18}\cline{20-20}
   &    &    &   &    &    &    &    &     &             & x & x & x  & x &\gx  &   & \gx  &   & &+   \\ \cline{11-18}\cline{20-20}
   &    &    &   &    &    &    &    &     &             & x &\gx&\gx& x & x  &    &  x   &    & &?  \\ \cline{11-18}\cline{20-20}
   &    &    &   &    &    &    &    &     &             & x & x &\gx & x & x  &    & \gx &   &  &+   \\ \cline{11-20}\cline{2-9}
   & x  & x  & x & x  &\gx &\gx &    &     &             & x & x &\gx&\gx&     &      &  x   &  x  & $q=4$&? \\ \cline{11-18}\cline{20-20}
   &    &    &   &    &    &    &    &     &             & x & x & x  &\gx&     &      & \gx &  x  &  &+   \\ \cline{11-18}\cline{20-20}
   &    &    &   &    &    &    &    &     &             & x & x & x  & x &     &   & \gx & \gx &  &+  \\ \cline{11-18}\cline{20-20}\cline{2-9}
   & x  & x &x   &\gx &\gx & x  &    &     &             & x & x &\gx &\gx&     &      &  x  &   x  &  &$-$   \\ \cline{11-18}\cline{20-20}
   &    &    &   &    &    &    &    &     &             & x & x & x  &\gx&     &      & \gx &  x  &  &+   \\ \cline{11-18}\cline{20-20}
   &    &    &   &    &    &    &    &     &             & x &\gx&\gx& x &     &      &  x  & x   &  &$-$     \\ \cline{11-18}\cline{20-20}
   &    &    &   &    &    &    &    &     &             & x & x  &  x & x &     &      & \gx&\gx & &+     \\ \cline{11-18}\cline{20-20}
   &    &    &   &    &    &    &    &     &             & x & x  &\gx&  x&   &    & \gx& x   &  &0    \\ \cline{11-18}\cline{20-20}\cline{2-9}
   &  x & x  &\gx&\gx &  x & x  &    &     &             & x & x  &\gx&\gx&     &      &  x & x   &   &$-$     \\ \cline{11-18}\cline{20-20}
   &    &    &   &    &    &    &    &     &             & x &\gx& x  &\gx&     &      &  x & x    &  &$-$     \\ \cline{11-18}\cline{20-20}
   &    &    &   &    &    &    &    &     &             & x & x  & x  &\gx&     &      &\gx& x    &  &$-$   \\ \cline{11-18}\cline{20-20}
   &    &    &   &    &    &    &    &     &             & \gx&\gx& x & x  &     &      & x  & x    &  &$-$    \\ \cline{11-18}\cline{20-20}
   &    &    &   &    &    &    &    &     &             & x  & x  & x  & x  &     &      &\gx&\gx & &?  \\ \cline{11-18}\cline{20-20}
   &    &    &   &    &    &    &    &     &             & x  &\gx & x & x   &    &    &\gx& x   &  &$-$   \\ \hline\hline
D6 & x  & x  &x  &\gx &\gx & x  &    &     &    D4       & x &\gx &\gx &   &     &      &  x &       &$q=3$ &? \\ \cline{11-18}\cline{20-20}
   &    &    &   &    &    &    &    &     &             & x & x &\gx &   &     &      &\gx &   &  &+ \\ \cline{11-18}\cline{20-20}\cline{2-9}
   & x  & x  &\gx&\gx & x  & x  &    &     &             & x  &\gx&\gx&   &     &      &  x  &       &  &$-$  \\ \cline{11-18}\cline{20-20}
   &    &    &   &    &    &    &    &     &             & \gx&\gx& x  &   &     &      &  x  &       &  &$-$  \\ \cline{11-18}\cline{20-20}
   &    &    &   &    &    &    &    &     &             &  x & x  &\gx&   &     &      &\gx &       & &+  \\ \cline{11-18}\cline{20-20}
   &    &    &   &    &    &    &    &     &             & x  &\gx & x  &   &     &   &\gx &   &  &0   \\ \cline{11-18}\cline{20-20}\cline{2-9}
   &  x &\gx &\gx& x  & x  &  x &    &     &             & x & \gx&\gx&   &     &      & \gx &       &  &$-$   \\ \cline{11-18}\cline{20-20}
   &    &    &   &    &    &    &    &     &             & \gx &\gx& x  &   &     &      &  x  &       &  &$-$   \\ \cline{11-18}\cline{20-20}
   &    &    &   &    &    &    &    &     &             & x  &\gx&  x &   &     &      &\gx &       & &$-$    \\ \cline{11-18}\cline{20-20}
   &    &    &   &    &    &    &    &     &             & \gx  & x  & x  &   &    &   &\gx &   & &$-$  \\ \cline{11-20}\cline{2-9}
   & x  & x  &\gx&\gx & x  & x  &    &     &             &\gx &\gx  &   &   &     &    & x  &  x   & $q=2$ &$-$   \\ \cline{11-18}\cline{20-20}
   &    &    &   &    &    &    &    &     &             & x  &\gx  &   &   &     &      &\gx  & x  &  &$-$   \\ \cline{11-18}\cline{20-20}
   &    &    &   &    &    &    &    &     &             & x  & x  &   &   &     &   &\gx  &\gx  &  &?\\ \cline{11-18}\cline{20-20}\cline{2-9}
   & x  &\gx &\gx& x  & x  & x  &    &     &             &\gx  &\gx  &   &   &     &      &  x  & x  & &$-$    \\ \cline{11-18}\cline{20-20}
   &    &    &   &    &    &    &    &     &             & \gx & x  &   &   &     &      &\gx  & x  &  &$-$   \\ \cline{11-18}\cline{20-20}
   &    &    &   &    &    &    &    &     &             & x  &\gx  &   &   &     &      &\gx & x &  &$-$   \\ \cline{11-18}\cline{20-20}
   &    &    &   &    &    &    &    &     &             & x  & x  &   &   &   &   &\gx &\gx & &$-$   \\ 
D6 &\gx &\gx & x & x  & x  & x  &    &     &    D4       & \gx & \gx &   &   &    &   &  x  &  x  &$q=2$ &$-$    \\ \cline{11-18}\cline{20-20}
   &    &    &   &    &    &    &    &     &             & x  &\gx  &   &   &     &      &\gx  & x  & &$-$    \\ \cline{11-18}\cline{20-20}
   &    &    &   &    &    &    &    &     &             & x  & x  &   &   &     &      & \gx & \gx&  &$-$   \\ \hline\hline
D6 & x  &\gx &\gx& x  & x  & x  &    &     &    D2       & \gx &   &   &  & &   & \gx  & & $q=1$ &$-$  \\ \cline{11-18}\cline{20-20}\cline{2-9}
   & \gx&\gx & x &x   & x  &  x &    &     &             & \gx &   &   &   &     &    & \gx &   & &$-$  \\ \cline{11-20}\cline{2-9}
   &\gx &\gx & x & x  & x  & x  &    &     &             &  &   &   &   &     &      &\gx &\gx  &$q=0$ &$-$  \\ \hline\hline
D4 & x  & x  &\gx&\gx &    &    &    &     &    D4       & x & \gx  & \gx &   &  x   &      &    &    &$q=3$  &+ \\ \cline{11-18}\cline{20-20}
   &    &    &   &    &    &    &    &     &             & x  & x  &  \gx &   &  \gx   &     &     &       & &+  \\ \cline{11-18}\cline{20-20}
   &    &    &   &    &    &    &    &     &             & \gx  & \gx  & x  &   &  x   &      &     &    &  &+   \\ \cline{11-18}\cline{20-20}
   &    &    &   &    &    &    &    &     &             & x & \gx & x  &   & \gx &    &   &    &  &+  \\ \cline{11-18}\cline{20-20}\cline{2-9}
   & x  &\gx &\gx& x  &    &    &    &     &             &  x & \gx & \gx &   & x &      &      &       &  &+   \\ \cline{11-18}\cline{20-20}
   &    &    &   &    &    &    &    &     &             & \gx & x  & \gx &   &  x   &      &      &     &  &+   \\ \cline{11-18}\cline{20-20}
   &    &    &   &    &    &    &    &     &             & x  & x  &\gx&   &  \gx &      &      &       &  &+   \\ \cline{11-18}\cline{20-20}
   &    &    &   &    &    &    &    &     &             & \gx  & x  & x  &   & \gx  &   &   &    &  &+ \\ \cline{11-20}\cline{2-9}
   & x  & x  &\gx&\gx &    &    &    &     &             &\gx&x&  &   & x &  x  &      &       &$q=2$ &?    \\ \cline{11-18}\cline{20-20}
   &    &    &   &    &    &    &    &     &             & \gx& x &   &   &\gx& x &      &       &  &+   \\ \cline{11-18}\cline{20-20}
   &    &    &   &    &    &    &    &     &             & x & x &   &   &\gx &\gx&      &     &  &+   \\ \cline{11-18}\cline{20-20}\cline{2-9}
   & x  &\gx &\gx& x  &    &    &    &     &             &\gx &\gx &   &   & x  &  x &      &       & &$-$    \\ \cline{11-18}\cline{20-20}
   &    &    &   &    &    &    &    &     &             & x &\gx &   &   &\gx &  x  &      &       & &+ \\ \cline{11-18}\cline{20-20}
   &    &    &   &    &    &    &    &     &             & \gx & x &   &   &\gx &x &      &       &   &0  \\ \cline{11-18}\cline{20-20}
   &    &    &   &    &    &    &    &     &             & x & x &   &   &\gx  &\gx &    &     &  &+   \\ \cline{11-18}\cline{20-20}\cline{2-9}
   &\gx &\gx & x &  x &    &    &    &     &             &\gx &\gx &   &   &  x  & x &      &       &  &$-$   \\ \cline{11-18}\cline{20-20}
   &    &    &   &    &    &    &    &     &             & x &\gx &   &   &\gx & x  &      &       &  &$-$   \\ \cline{11-18}\cline{20-20}
   &    &    &   &    &    &    &    &     &             & x & x &   &   &\gx &\gx &      &     & &?  \\ \cline{11-20}\cline{2-9}
   & x  &\gx &\gx&x   &    &    &    &     &             &\gx &  &   &   &\gx & x  & x &       &$q=1$ &$-$    \\ \cline{11-18}\cline{20-20}
   &    &    &   &    &    &    &    &     &             & x &   &   &   &\gx  &\gx & x  &    &  &?   \\ \cline{11-18}\cline{20-20}\cline{2-9}
   &\gx & \gx& x & x  &    &    &    &     &             & \gx  &   &   &   &\gx  & x  & x &       &  &$-$   \\\cline{11-18}\cline{20-20}
   &    &   &    &    &    &    &    &     &             & x  &   &   &   &\gx & \gx &  x &     &  &$-$ \\ \cline{11-20}\cline{2-9}
   &\gx &\gx& x  & x  &    &    &    &     &             &   &   &   &   &\gx &\gx  & x  &  x  & $q=0$ &$-$   \\ \hline\hline
D4 & x  &\gx&\gx & x  &    &    &    &     &     D2      & \gx  &   &   &   & \gx &   &  &  &$q=1$ &+  \\ \cline{11-18}\cline{20-20}\cline{2-9}
   &\gx &\gx& x  & x  &    &    &    &     &             &\gx  &   &   &   & \gx   &    &   &   &  &+   \\ \cline{11-20}\cline{2-9}
   &\gx &\gx& x  & x  &    &    &    &     &             &   &   &   &   & \gx  & \gx  &      &       &$q=0$  &?   \\\hline\hline
D2 &\gx &\gx&    &    &    &    &    &     &     D2      & \gx &   &\gx&   &    &    &    &    & $q=1$ &+   \\ \cline{11-20}\cline{2-9}
   &\gx &\gx&    &    &    &    &    &     &             &   &   &\gx&\gx &     &      &      &     & $q=0$ &+   \\ \hline
\end{longtable}

\end{document}